%% file: main.tex
\documentclass[11pt,draftcls,onecolumn]{IEEEtran}
\ifCLASSINFOpdf
\else
   \usepackage[dvips]{graphicx}
\fi
\usepackage{url}
\usepackage{graphicx}
\usepackage{amsmath}
\usepackage{comment}
\usepackage[utf8]{inputenc}
\usepackage[english]{babel}
\usepackage{color}
\usepackage{framed}
\usepackage{amsmath,amssymb,amsfonts}

\usepackage{graphicx}
\usepackage{upgreek}
\usepackage{textcomp}

\usepackage{url}

\usepackage{cite}

\input{symbols.tex}
\usepackage{bbold}
\usepackage{ifpdf}
\usepackage{times}
\usepackage{layout}
\usepackage{float}
\usepackage{afterpage}
\usepackage{amsmath}
\usepackage{amstext}
\usepackage{amssymb,bm}
\usepackage{stmaryrd}
\usepackage{latexsym}
\usepackage{color}
\usepackage{flexisym}
\usepackage{kantlipsum}
\usepackage{graphicx}
\usepackage{amsmath}
\usepackage{amsthm}
\usepackage{graphicx}

\usepackage[caption=false,font=footnotesize]{subfig}

\usepackage{booktabs}
\usepackage{multicol}

\usepackage{enumitem}

\usepackage[switch]{lineno}
\usepackage[named]{algo}
\usepackage[noend]{algpseudocode}
\usepackage{algorithm}

\makeatletter
\newcommand*{\rom}[1]{\expandafter\@slowromancap\romannumeral #1@}
\makeatother

\begin{document}
\setlength{\abovedisplayskip}{3pt}
\setlength{\belowdisplayskip}{3pt}

\title{HDR Imaging With One-Bit Quantization
}

\author{Arian Eamaz, \IEEEmembership{Graduate Student Member, IEEE}, Farhang Yeganegi, and Mojtaba Soltanalian, \IEEEmembership{Senior Member, IEEE}

\vspace{-20pt}
\thanks{This work was supported in part by the National Science Foundation Grant CCF-1704401. (\emph{Corresponding author: Arian Eamaz}). The first two authors contributed equally to this work.}
\thanks{The authors are with the Department of Electrical and Computer Engineering, University of Illinois Chicago, Chicago, IL 60607, USA (e-mail: \emph{ \{aeamaz2, fyegan2, msol\}@uic.edu}).}
}
\markboth{
}
{Shell \MakeLowercase{\textit{et al.}}: Bare Demo of IEEEtran.cls for IEEE Journals}
\maketitle

\begin{abstract}
Modulo sampling and dithered one-bit quantization frameworks have emerged as promising solutions to overcome the limitations of traditional analog-to-digital converters (ADCs) and sensors. Modulo sampling, with its high-resolution approach utilizing modulo ADCs, offers an unlimited dynamic range, while dithered one-bit quantization offers cost-efficiency and reduced power consumption while operating at elevated sampling rates. Our goal is to explore the synergies between these two techniques, leveraging their unique advantages, and to apply them to non-bandlimited signals within spline spaces. One noteworthy application of these signals lies in High Dynamic Range (HDR) imaging.
In this paper, we expand upon the Unlimited One-Bit (UNO) sampling framework, initially conceived for bandlimited signals as proposed in \cite{eamaz2022uno}, to encompass non-bandlimited signals found in the context of HDR imaging. 
We present a novel algorithm rigorously examined for its ability to recover images from one-bit modulo samples. Additionally, we introduce a sufficient condition specifically designed for UNO sampling to perfectly recover non-bandlimited signals within spline spaces. Our numerical results vividly demonstrate the effectiveness of UNO sampling in the realm of HDR imaging.
\end{abstract}

\begin{IEEEkeywords}
HDR imaging, modulo sampling, non-bandlimited signals, one-bit quantization, UNO sampling. 
\end{IEEEkeywords}

\setlength{\abovedisplayskip}{3pt}
\setlength{\belowdisplayskip}{3pt}

\section{Introduction}
\label{intro}
High Dynamic Range (HDR) imaging, a pivotal domain within image processing, computer graphics, vision, and photography, revolutionizes traditional imaging by enabling a vastly extended range of exposure. Its primary objective is to faithfully depict the full spectrum of intensity levels present in real-world scenes, spanning from intense sunlight to subtle shadows \cite{banterle2017advanced,tiwari2015review}. 
HDR imaging allows the faithful capture, storage, transmission, and utilization of real-world lighting in a variety of applications, all without the need for signal linearization or managing clipped values \cite{tiwari2015review}. 
Due to sensor saturation, when the signal's dynamic range exceeds a certain threshold, denoted as $\lambda$, it results in localized information loss. Traditionally, specialized HDR cameras were the primary means of capturing HDR images. Nevertheless, these cameras are so expensive \cite{johnson2015high,wang2017angular}. 
To address the issue of saturation in HDR imaging, one potential solution is to implement the approach outlined in \cite{bhandari2020hdr}. The upshot of \cite{bhandari2020hdr} is that they provided a method for single-shot HDR imaging that exploits signal-dependent quantization noise. They characterized the quantization noise by folding the input signal values within the range of $-\lambda$ to $\lambda$, effectively transforming them into modulo samples. These modulo samples were then used as key information for reconstructing the signal. This innovative approach allowed them to capture and utilize quantization noise and amplitudes that exceeded the sensor's limitations.
Remarkably, this method capitalizes on the inherent spatial anti-aliasing performed by all imaging sensors \cite{lukac2018single}. Unlike many other techniques for recovering images from sensor data, this apparent constraint actually works to their advantage. 
The smoothness introduced by spatial anti-aliasing \cite{lukac2018single} imbues the quantization noise with a structured pattern of modulo samples, which their approach exploits during the recovery process \cite{bhandari2020hdr}.

The concept of modulo sampling, aimed at mitigating information loss caused by signal clipping in conventional ADCs, has been extensively studied in existing literature \cite{rhee2003wide,boufounos2011universal,bhandari2017unlimited,ordentlich2018modulo}. In a related context, \cite{rhee2003wide} introduced an HDR imaging sensor utilizing self-reset ADCs that either reset or fold when signal amplitudes reach the saturation threshold. These ADC designs, though, capture both modulo samples and a count of the number of resets, both of which are subsequently employed in tandem for signal reconstruction.
Conversely, the approach of \emph{unlimited sampling}, as proposed in \cite{bhandari2017unlimited,bhandari2020unlimited}, utilizes modulo ADCs but exclusively relies on the modulo samples for reconstructing the input signal. More precisely, a modulo operator applied to input signal $g(t)$ is defined as

\begin{equation}
\mathcal{M}_\lambda: g \mapsto 2 \lambda\left(\left\llbracket \frac{g}{2 \lambda}+\frac{1}{2} \right\rrbracket-\frac{1}{2}\right),\quad \llbracket g \rrbracket  \triangleq g-\lfloor g\rfloor, \quad \lambda \in \mathbb{R}^{+}.
\end{equation}

Conventional multi-bit ADCs demand a substantial number of quantization levels to accurately represent continuous signals at high resolutions. However, employing high-resolution ADCs at elevated data rates significantly amplifies both power consumption and manufacturing costs \cite{wagdy1989validity}. This challenge becomes even more pronounced in systems requiring multiple ADCs, as seen in extensive array receivers \cite{ho2019antithetic}. To address these issues, recent efforts have centered on the development of receivers featuring low-complexity, one-bit ADCs. In the realm of one-bit quantization, ADCs essentially compare incoming signals against predetermined threshold levels, yielding binary outputs of either $+1$ or $-1$.
This approach permits high-rate signal sampling while maintaining substantially lower costs and energy consumption compared to traditional ADCs\cite{eamaz2021modified}. Numerous applications benefit from one-bit ADCs, including multiple-input multiple-output wireless communications \cite{mezghani2018blind}, and array signal processing \cite{liu2017one}. 
The literature extensively demonstrates that signal acquisition using coarse quantization with time-varying sampling thresholds (random dithering) yields markedly superior results compared to both zero and constant threshold approaches \cite{dirksen2022covariance,eamaz2022covariance,eamaz2022phase,xu2020quantized}.
Significant research has explored one-bit sigma-delta quantization, investigating the influence of adaptive threshold selection in achieving rapid reduction of reconstruction errors concerning the number of measurements or oversampling rates, depending on the specific scenario. For more in-depth insights, interested readers can refer to relevant works such as \cite{gunturk2003one,deift2011optimal}. 
\subsection{Motivation and Contributions of The Paper}
In this paper, our motivation stems from \cite{bhandari2020hdr}, driving our exploration into the impact of coarse quantization on HDR imaging when applying dithered one-bit quantization to modulo samples. Specifically, we investigate the feasibility of reconstructing a signal using only one-bit data instead of storing the modulo samples. The concept of combining unlimited sampling and coarse quantization was initially introduced by \cite{graf2019one,eamaz2022uno}.
In \cite{eamaz2022uno}, the authors proposed the unlimited one-bit (UNO) sampling leveraging modulo sampling to design a dithering scheme for coarse quantization. This scheme tailors the statistical properties of thresholds based on information from the ADC threshold value, denoted as $\lambda$, to align the dynamic range of modulo samples and dithers. They both numerically and theoretically showcased that aligning the modulo samples and dithers results in more informative one-bit samples, ultimately leading to enhanced signal reconstruction performance for \emph{bandlimited signals}.
Traditionally, we do not possess information about the dynamic range of the input signal. However, with the implementation of modulo sampling, we gain insight into the modulo samples' dynamic range. This knowledge enables us to craft a judicious dithering scheme that closely approximates the signal values and preserves information about the distances between the signal and thresholds. Furthermore, studies have indicated that the signal reconstruction performance achieved through dithered one-bit quantization with unlimited sampling surpasses that of sigma-delta quantization \cite{eamaz2022uno}. Note that the application of one-bit sigma-delta quantization in HDR imaging appears impractical due to its inherent feedback loop and non-ideal noise shaping characteristics. These factors can introduce complexities in mathematical analysis and the sensing process\cite{graf2019one}.

We extend our focus beyond bandlimited input signals and delve into \emph{assessing the performance of UNO sampling in reconstructing signals that belong to spline spaces (non-bandlimited), a prevalent model in various imaging applications.} Our approach will involve a thorough examination of the sampling theorem, and we will introduce a randomized algorithm designed for the comprehensive reconstruction of input images, accompanied by a rigorous convergence analysis. We extend the unlimited sampling theorem, which provides sufficient conditions for perfect reconstruction, to encompass scenarios where the modulo samples may be corrupted. Subsequently, we derive the UNO sampling theorem based on this extended framework. \emph{Our numerical results demonstrate that simplifying the imaging system by employing one-bit ADCs, which can be as basic as a simple comparator, still yields an effective imaging process with high-quality reconstructed images.}

\textbf{Notation:} Throughout this paper, we use boldface lowercase, boldface uppercase, and calligraphic letters for vectors, matrices, and sets, respectively. The notations $\mathbb{C}$, $\mathbb{R}$, and $\mathbb{Z}$ represent the set of complex, real, and integer numbers, respectively. For a vector $\mathbf{x}$, $\Delta\mathbf{x}=x_{k+1}-x_{k}$ denotes the finite difference and recursively applying the same yields $N$-th order difference, $\Delta^{N}\mathbf{x}$. The infinity or max-norm 
of a function $x$ is  $\|x\|_{\infty}=\operatorname{inf}\left\{c_{0}\geq 0: |x(t)|\leq c_{0}\right\}$, for vectors, we have $\|\mathbf{x}\|_{\infty}=\max_{k}|x_{k}|$. The function $\operatorname{sgn}(\cdot)$ yields the sign of its argument. 
In the context of numerical computations, $\lfloor\rfloor$ denotes the floor function. Given a scalar $x$, we define the operator $(x)^{+}$ as $\max\left\{x,0\right\}$. A ball with radius $r$ centered at a point $\mby\in\mathbb{R}^{n}$ is defined as $\mathcal{B}_r\left(\mby\right)=\left\{\mby_1\in\mathbb{R}^{n}|\left\|\mby-\mby_1\right\|_2\leq r\right\}$. The set $[n]$ is defined as $[n]=\left\{1,\cdots,n\right\}$.

\section{Problem Setup and Sampling Pipeline}
\label{OB-MC}
Consider $\mathbf{x} \in \mathbb{R}^d$ as the spatial coordinates within a $d$-dimensional image denoted as $f(\mathbf{x})$. When capturing such an image using a digital imaging sensor, it undergoes spatial pre-filtering through two distinct stages.
Initially, pixels do not go through point-wise sampling; instead, they accumulate incoming light over a finite area, a process known as ``average sampling." This leads to spatial smoothing at the pixel level. Additionally, this effect becomes prominent in color sensors employing a grid of colored elements, often referred to as the Bayer Pattern, responsible for generating colored images. Pixel-level pre-filtering and light sensitivity are typically enhanced using micro-lens arrays. A secondary layer of pre-filtering comes into play for spatial anti-aliasing \cite{lukac2018single}, smoothing out finer details beyond the sensor's spatial resolution.
Modern sensors are meticulously optimized to minimize the loss of details resulting from inevitable filtering. These effects are subsequently compensated for in later stages of the imaging pipeline during processes such as colorization and sharpening of digitized images. The techniques presented in this paper are specifically designed to intervene in the sensing pipeline before these image processing mechanisms are applied.

In order to achieve a mathematical approximation of anti-aliased natural images, which exhibit smoothness at extremely fine scales while maintaining accurate representation even at finer scales, we will utilize shift-invariant spline expansion \cite{unser1999splines}. The approximated image by splines $g(\mbx)$ is given by

\begin{equation}
\label{Steph1}
g(\mbx)=\sum_{\mathbf{m} \in \mathbb{Z}^d} c[\mathbf{k}] \mathrm{B}_N\left(\mathbf{H}^{-1} \mbx-\mathbf{m}\right)
\end{equation}
where $\mathbf{H}$ is a $d \times d$ diagonal matrix, with dilations $h_1, h_2, \ldots, h_d$ on the diagonal, $\mathbf{c}$ are the coefficients and $\mathrm{B}_N$ is a B-spline of or $\operatorname{der} N$. Since B-spline bases are separable, the tensor product representation holds, $\mathrm{B}_N(\mathbf{x})=$ $\prod_{m=1}^d \mathrm{~B}_N\left(x_m\right)$. For simplicity of exposition, we will work with $d=1$  and use separable filtering. The coefficients in \eqref{Steph1}, can be evaluated using different strategies, comprehensively discussed in \cite{blu1999quantitative}. It was demonstrated in \cite{blu1999quantitative} that for $f \in \mathbf{L}_2$, its approximation via interpolation $g$, entails an error bound, $\|f-g\|_{2} \propto h^L\left\|g^{(L)}\right\|_{2}, h \rightarrow 0$ where $L=N+1$ is the approximation order, $g^{(L)}$ is the derivative of order $L$. In the UNO sampling framework, we firstly apply the modulo operator to samples of $g$
\begin{equation}
 \label{Steph2}   
 y_k = \mathcal{M}_{\lambda}\left(g\left(k T\right)\right),
\end{equation}
where $1/T$ is sampling rate. Then, we generate one-bit data with one-bit quantization using multiple dithering sequences $\left\{\boldsymbol{\uptau}^{(\ell)}\right\}^{m}_{\ell=1}$, as follows
\begin{equation}
\mbr^{(\ell)} = \operatorname{sgn}\left(\mby-\boldsymbol{\uptau}^{(\ell)}\right), \quad \ell\in [m],
\end{equation}
or equivalently, 
\begin{equation}
\label{Steph4}
\mbr^{(\ell)} \odot \left(\mby-\boldsymbol{\uptau}^{(\ell)}\right) \succeq \mathbf{0},\quad \ell\in [m].    
\end{equation}
It follows from (\ref{Steph4}) that 
\begin{equation}
\label{eq:7}
\begin{aligned}
\bOmega^{(\ell)} \mby &\succeq \mathbf{r}^{(\ell)} \odot \boldsymbol{\uptau}^{(\ell)}, \quad \ell \in [m],
\end{aligned}
\end{equation}
where the sign matrix $\bOmega^{(\ell)}=\operatorname{diag}\left(\mathbf{r}^{(\ell)}\right)$. Denote the concatenation of all $m$ sign matrices as 
\begin{equation}
\label{eq:9}
\Tilde{\bOmega}=\left[\begin{array}{c|c|c}
\bOmega^{(1)} &\cdots &\bOmega^{(m)}
\end{array}\right]^{\top}, \quad \Tilde{\bOmega}\in \mathbb{R}^{m n\times n}.
\end{equation}
Rewrite the $m$ linear 
inequalities in \eqref{eq:7} as 
\begin{equation}
\label{eq:8}
\Tilde{\bOmega} \mby \succeq \operatorname{vec}\left(\mathbf{R}\right)\odot \operatorname{vec}\left(\bGamma\right),
\end{equation}
where $\mathbf{R}$ and $\bGamma$ are matrices, whose columns are the sequences $\left\{\mathbf{r}^{(\ell)}\right\}_{\ell=1}^{m}$ and $\left\{\boldsymbol{\uptau}^{(\ell)}\right\}_{\ell=1}^{m}$, respectively. The linear system of inequalities in (\ref{eq:8}) associated with the one-bit sampling scheme is overdetermined. We recast (\ref{eq:8}) into a \textit{one-bit polyhedron} as
\begin{equation}
\label{eq:8n}
\begin{aligned}
\mathcal{P}_{\mby} = \left\{\mby \mid \Tilde{\bOmega} \mby \succeq \operatorname{vec}\left(\mathbf{R}\right)\odot \operatorname{vec}\left(\bGamma\right)\right\}.
\end{aligned}
\end{equation}
Note that this shrinking space always contains the global minima, with a
volume that is diminished with an increased sample size. 

With the incorporation of modulo sampling, we gain insight into the DR of $\mby$, which is equal to $\operatorname{DR}_{\mby}=\lambda$. This information enables us to craft a dithering scheme wherein the threshold values closely align with the DR of modulo samples.
Given that the modulo samples are uniformly distributed within the range of $-\lambda$ to $\lambda$, we opt for a Uniform dithering scheme where thresholds are generated as $\boldsymbol{\uptau}^{(\ell)}\sim\mathcal{U}_{\left[-\lambda,\lambda\right]}$. This enables storing the information on the distance
between the modulo samples and the thresholds without any loss of
information via one-bit sensing. The implementation of a dithered generator, specifically for generating Gaussian and Uniform dithering in quantization systems within an ADC system, was described in \cite{robinson2019analog}.  Additionally, the implementation of multiple random dithering was demonstrated in \cite{ali2020background} for a 12-bit, 18-GS/s ADC.

\section{Recovery From One-Bit Modulo Samples}
\label{sec3}

Our proposed algorithm for reconstructing the original signal from one-bit data follows these sequential steps:
\begin{enumerate}
    \item Reconstruct the modulo samples $\mby\in \mathbb{R}^{n}$ from one-bit data $\left\{\mbr^{(\ell)}\right\}^{m}_{\ell=1}$ by solving the inequality system formed in (\ref{eq:8n}). 
    \item Reconstruct the original image from modulo samples by employing the algorithm proposed in \cite{bhandari2020hdr}. 
\end{enumerate}
To obtain the modulo samples $\mby$ in the polyhedron (\ref{eq:8n}), it is required to solve a linear system of inequalities. We tackle this polyhedron search problem through the randomized Kaczmarz algorithm (RKA) because of its optimal randomized projection and linear convergence in expectation \cite{leventhal2010randomized}. One intriguing advantage of RKA lies in its utilization of only a single measurement at each iteration, owing to its randomized nature. 
The RKA is a \emph{sub-conjugate gradient method} to solve a linear feasibility problem, i.e, $\Tilde{\bOmega}\mby\succeq\mathbf{b}=\operatorname{vec}\left(\mathbf{R}\right)\odot \operatorname{vec}\left(\bGamma\right)$ where $\Tilde{\bOmega}$ is a ${m n\times n}$ matrix. Conjugate-gradient methods immediately turn the mentioned inequality to an equality in the following form $\left(\mbb-\Tilde{\bOmega}\mby\right)^{+}=0$,
and then, approach the solution by the same process as used for systems of equations \cite{leventhal2010randomized}.
The projection coefficient $\beta_{i}$ of the RKA is \cite{leventhal2010randomized}
\begin{equation}
\label{eq:22}
\beta_{i}= \begin{cases}
\left(b_{j}-\mathbf{c}_{j}\mby_{i}\right)^{+} & \left(j \in \mathcal{I}_{\geq}\right), \\ b_{j}-\mathbf{c}_{j} \mby_{i} & \left(j \in \mathcal{I}_{=}\right),
\end{cases}
\end{equation}
where the disjoint index sets $\mathcal{I}_{\geq}$ and $\mathcal{I}_{=}$ partition $\mathcal{J}$ and $\{\mathbf{c}_{j}\}$ are the rows of $\Tilde{\bOmega}$.
Also, the unknown column vector $\mby$ is iteratively updated as
\begin{equation}
\label{eq:23}
\mby_{i+1}=\mby_{i}+\frac{\beta_{i}}{\left\|\mbc_{j}\right\|^{2}_{2}} \mbc^{\mathrm{H}}_{j},
\end{equation}
where, at each iteration $i$, the index $j$ is drawn from the set $\mathcal{J}$ independently at random following the distribution
$\operatorname{Pr}\{j=k\}=\frac{\left\|\mbc_{k}\right\|^{2}_{2}}{\|\Tilde{\bOmega}\|_{\mathrm{F}}^{2}}$.
Assuming that the linear system is consistent with nonempty feasible set $\mathcal{P}_{\mby}$ created by the intersection
of hyperplanes around the desired point $\mby_{\star}$,
RKA converges linearly in expectation to the solution $\hat{\mby}\in\mathcal{P}_{\mby}$:
\begin{equation}
\label{eq:15}
\mathbb{E}\left\{\left\|\mby_{i}-\hat{\mby}\right\|_{2}^{2}\right\} \leq \left(1-q_{_{\text{RKA}}}\right)^{i}~ \left\|\mby_{0}-\hat{\mby}\right\|_{2}^{2},
\end{equation}
where 
$i$ is the number of required iterations for RKA, and $q_{_{\text{RKA}}} \in \left(0,1\right)$ is given by $q_{_{\text{RKA}}}=\frac{1}{\kappa^{2}\left(\Tilde{\bOmega}\right)}$,
with $\kappa\left(\Tilde{\bOmega}\right)=\|\Tilde{\bOmega}\|_{\mathrm{F}}\|\Tilde{\bOmega}^{\dagger}\|_{2}$ denoting the scaled condition number of a matrix $\Tilde{\bOmega}$. As shown in \cite{eamaz2023harnessing}, $\kappa^{2}$ of one-bit data matrix is equal to $m$. 

The convergence analysis of RKA demonstrates its ability to converge to a solution within the feasible set of hyperplanes. However, if our objective is to converge to a solution within a specified error radius, denoted as $\rho$, around the original signal, the convergence rate becomes contingent on the quantity of one-bit samples used. This limitation implies that it does not account for the impact of increasing the number of samples and the proximity of the solution to the original signal when our goal is a solution within a specific error radius.
To elucidate the relationship between the number of samples ($m^{\prime}=m n$) and the convergence rate of RKA to the original signal, we propose the following convergence rate:
\begin{proposition}
\label{Stephanie_proposition}
Consider the one-bit polyhedron $\mathcal{P}_{\mby}$ obtained in \eqref{eq:8n} associated with the dithered one-bit sensing \eqref{Steph4} and $\delta$ as a positive constant. 
Consider a ball centered at the original modulo signal $\mathcal{B}_{\rho}\left(\mby_{\star}\right)$ and $\hat{\mby}\in\mathcal{P}_{\mby}$, a convergence rate for RKA may be formulated as:
\begin{equation}
\label{bound2}
\begin{aligned}
\mathbb{E}\left\{\left\|\mby_{i}-\mby_{\star}\right\|_{2}^{2}\right\} \leq \left(1-\frac{1}{m}\right)^{i} \left\|\mby_{0}-\hat{\mby}\right\|_{2}^{2}+\rho^2,
\end{aligned}
\end{equation}
with a probability exceeding $1-e^{-\delta m^{\prime}}$.
\end{proposition}
\begin{IEEEproof}
As demonstrated in \cite{leventhal2010randomized}, the solution obtained from RKA lies in the space formed by the hyperplanes of the linear inequality problem with the following convergence rate:
\begin{equation}
\label{bound200}
\begin{aligned}
\mathbb{E}\left\{\left\|\mby_{i}-\hat{\mby}\right\|_{2}^{2}\right\} \leq \left(1-\frac{1}{m}\right)^{i} \left\|\mby_{0}-\hat{\mby}\right\|_{2}^{2},
\end{aligned}
\end{equation}
where $\hat{\mby}$ is a point inside the space created by a polyhedron $\mathcal{P}_{\mby}$. The convergence rate \eqref{bound200} only ensures that the solution will lie within the space created by the hyperplanes, not necessarily within the ball around the desired solution. However, in order to guarantee a perfect reconstruction and ensure that the solution of the linear feasibility problem lies within the ball around the desired solution with radius $\rho$, it is essential to have a sufficient number of samples. Until we reach the required number of samples, the upper bound of convergence to a solution lying within the space $\mathcal{P}_{\mby}$, differs (and is smaller than) from that of the convergence to a solution within the ball around the desired point, since $\operatorname{vol}\left(\mathcal{P}_{\mby}\right) \nsubseteq \operatorname{vol}\left(\mathcal{B}_{\rho}(\mby_{\star})\right)$. However, once we obtain a sufficient number of samples to have the volume created by the intersections of hyperplanes inside the ball, i.e., $\operatorname{vol}\left(\mathcal{P}_{\mby}\right) \subseteq \operatorname{vol}\left(\mathcal{B}_{\rho}(\mby_{\star})\right)$, we then have $\hat{\mby}$  lying within the ball around the desired point $\mbx_{\star}$, i.e., $\hat{\mby}\in\mathcal{B}_{\rho}\left(\mby_{\star}\right)$. To address this discrepancy between the two scenarios, we introduce a second term that is dependent on the difference between $\hat{\mby}$ and $\mby_{\star}$, as follows:
\begin{equation}
\begin{aligned}
\mathbb{E}\left\{\left\|\mby_i-\mby_{\star}\right\|^2_{2}\right\} &= \mathbb{E}\left\{\left\|\mby_i-\hat{\mby}+\hat{\mby}-\mby_{\star}\right\|^2_{2}\right\}\\&\leq \mathbb{E}\left\{\left\|\mby_i-\hat{\mby}\right\|^2_{2}\right\} + \mathbb{E}\left\{\left\|\mby_{\star}-\hat{\mby}\right\|^2_{2}\right\},
\end{aligned}   
\end{equation}
where from  \eqref{bound200} and the fact that the error between $\hat{\mby}$ and the original modulo signal remains deterministic with respect to each iteration, we can write
\begin{equation}
\begin{aligned}
\mathbb{E}\left\{\left\|\mby_i-\mby_{\star}\right\|^2_{2}\right\} \leq \left(1-\frac{1}{m}\right)^{i} \left\|\mby_0-\hat{\mby}\right\|^2_{2} + \left\|\mby_{\star}-\hat{\mby}\right\|^2_{2}.
\end{aligned}
\end{equation}
The convergence to $\mby_{\star}$ is ensured only when the second term, $\left\|\mby_{\star}-\hat{\mby}\right\|^2_{2}$, is bounded. As comprehensively investigated in \cite[Theorem~3]{eamaz2023harnessing}, with a minimum probability of $1-e^{-\delta m^{\prime}}$, we establish that $\operatorname{vol}\left(\mathcal{P}_{\mby}\right) \subseteq \operatorname{vol}\left(\mathcal{B}_{\rho}(\mby_{\star})\right)$ and consequently have $\left\|\mby_{\star}-\hat{\mby}\right\|^2_{2} \leq \rho^2$, which proves the proposition. 
\end{IEEEproof}
Let $\mathcal{K}_N$ represent the Bohr-Farvard constant as detailed in \cite{bhandari2020hdr}. Furthermore, let $l\leq N$ denote the order of differences necessary for the reconstruction algorithm from modulo samples, which follows Itoh's method. Additionally, consider the parameter $h$ as an indicator of the spline interpolation applied to the original image, denoted as $f$. In the context of UNO sampling for HDR imaging (or any non-bandlimited signal expanded within spline spaces), the sampling rate is established by the following theorem:
\begin{theorem}
\label{Steph_theroem}
Define $\mathcal{E}_{(l, N)}$ as $\frac{\mathcal{K}_{N-l}}{\mathcal{K}_N}$, $\beta_g$ obtained as $\left\|g\right\|_{\infty} \leq \beta_g$, and both $h$ and $\delta^{\prime}$ as positive constants. The sufficient condition for the UNO reconstruction algorithm to effectively approximate the samples $\gamma_{k} = g(kT)$, up to additive multiples of $2\lambda$ is:
\begin{equation}
\label{Steph5}
T<\frac{h}{\pi e}\left(\frac{\lambda}{2\beta_g \mathcal{E}_{(l, N)}}\right)^{1 / l},\quad l \leqslant N,
\end{equation}
with a probability exceeding $1-e^{-\delta^{\prime} m^{\prime}}$.
\end{theorem}
\begin{IEEEproof}
As comprehensively studied in \cite{bhandari2020hdr}, the sufficient condition to reconstruct the signal from modulo samples is given by 
\begin{equation}
\label{Steph6}
T<\frac{h}{\pi e}\left(\frac{\lambda}{\beta_g \mathcal{E}_{(l, N)}}\right)^{1 / l},\quad l \leqslant N.
\end{equation}
The sufficient condition \eqref{Steph6} indicates that the $l$-difference of samples, denoted as $\Delta^{l} \boldsymbol{\gamma}$, must be constrained within $\lambda$. 
Achieving this bounded difference is essential as it enables us to eliminate the nonlinearity introduced by the modulo operator in the difference values. Subsequently, we can effectively apply the anti-difference operator to $\Delta^{l} \boldsymbol{\gamma}$ in order to reconstruct the samples. The upper bound of $\Delta^{l} \boldsymbol{\gamma}$ is provided in \cite{bhandari2020hdr} as follows:
\begin{equation}
\label{Steph7}
\left\|\Delta^l \boldsymbol{\gamma}\right\|_{\infty} \leqslant\left(\frac{T \pi \mathrm{e}}{h}\right)^l\left(\frac{\mathcal{K}_{N-l}}{\mathcal{K}_N}\right)\|g\|_{\infty}.
\end{equation}
Nevertheless, when we retrieve modulo samples from one-bit data, the resulting reconstructed modulo samples exhibit disparities from their original values, as indicated by the errors in the one-bit signal reconstruction algorithm. In essence, within the UNO scheme, our modulo samples are subject to noise corruption stemming from the impact of the one-bit signal reconstruction process. It appears that sufficient condition \eqref{Steph6} may not provide adequate assurance for the boundedness of $\Delta^{l} \boldsymbol{\gamma}$ in the presence of these corrupted modulo samples. More precisely, we assume that the modulo
samples $y_k$ are affected by noise $\mbe$ (RKA reconstruction error) bounded by a
constant $b_0$. That is,
$\Tilde{y}_k = y_k+e_k,\quad \left\|\mbe\right\|_{\infty} \leq b_0$.
Note that due the presence of error, it may happen that 
$\Tilde{y}_k \notin[-\lambda, \lambda]$.
Nonetheless, for $b_0$ below some fixed
threshold, our recovery method provably recovers noisy samples $\gamma_k$ from the associated noisy modulo
samples $\Tilde{\gamma_k}$ up to an unknown additive constant, where
the noise appearing in the recovered samples is in entry-wise
agreement with the one affecting the modulo samples. That is, $\Tilde{\gamma}_k=\gamma_k+e_k+2 m \lambda, m \in \mathbb{Z}$. In order to ensure that $\Delta^{l} \boldsymbol{\Tilde{\gamma}}$ remains within the bounds of $\lambda$, it is necessary for both $\Delta^{l} \boldsymbol{\gamma}$ and $\Delta^{l} \mbe$ to be constrained to $\lambda/2$. By considering both \eqref{Steph7} and $\left\|\Delta^{l}  \boldsymbol{\gamma}\right\|_{\infty}\leq \lambda/2$, one can easily obtain \eqref{Steph6}. On the other hand, according to Young's inequality \cite{bhandari2020unlimited} and $\left\|\Delta^{l}\mbe\right\|_{\infty}\leq \lambda/2$, we have
\begin{equation}
\label{Steph8}   
\left\|\Delta^{l}\mbe\right\|_{\infty}\leq 2^{l}\left\|\mbe\right\|_{\infty} \leq \frac{\lambda}{2},
\end{equation}
or equivalently   
$\left\|\mbe\right\|_{\infty} \leq \frac{\lambda}{2^{l+1}}$.
It means in the one-bit signal reconstruction, the following error bound must be met:
\begin{equation}
\left\|\mby_{\star}-\hat{\mby}\right\|_{\infty}\leq \frac{\lambda}{2^{l+1}}.
\end{equation}
Hence, considering that the infinity norm is consistently smaller than the second norm, it appears sufficient to determine the minimum probability at which the solution of the one-bit signal reconstruction falls within the ball $\mathcal{B}_{\rho}\left(\mby_{\star}\right)$, i.e., $\hat{\mby}\in\mathcal{B}_{\rho}\left(\mby_{\star}\right)$ with $\rho=\lambda/2^{l+1}$. As outlined in \cite{eamaz2023harnessing}, the minimum probability for a solution in one-bit signal reconstruction to fall within a ball of radius $\rho$ centered on the original signal is given by $1-e^{-\delta^{\prime} m^{\prime}}$. Here, it is important to note that $\delta^{\prime}$ is a monotonically increasing function in relation to $\rho$, and in this case, $\rho$ is selected as $\lambda/2^{l+1}$—a choice that serves to establish the theorem.
\end{IEEEproof}
By increasing the
number of time-varying threshold sequences in accordance
with the computed minimum probability, the likelihood of
achieving perfect reconstruction with the UNO sampling rate
\eqref{Steph5} also increases. Note that oversampling is a common scenario in one-bit
quantization techniques and is not a major concern in UNO
implementation.
\begin{figure}[t]
	\centering
    \subfloat[]
		{\includegraphics[width=0.3\columnwidth]{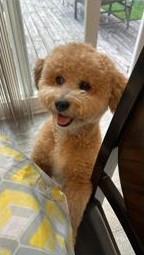}}\qquad 
    \subfloat[]
		{\includegraphics[width=0.3\columnwidth]{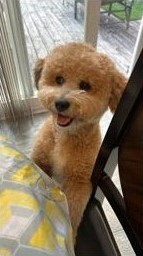}}\qquad
   \subfloat[]
		{\includegraphics[width=0.3\columnwidth]{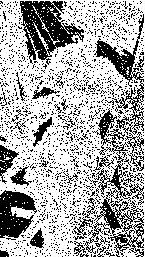}}
	\caption{(a) shows the original image of Grape (pet's name). (b) illustrates the reconstructed image from the UNO samples. As evident, the resulting image showcases high quality, devoid of any noticeable reconstruction artifacts discernible to the naked eye. (c) shows the UNO data for an arbitrary sequence of threshold.}
	\label{figure_1}
\end{figure}
\begin{figure}[t]
	\centering
	\includegraphics[width=0.6\columnwidth]{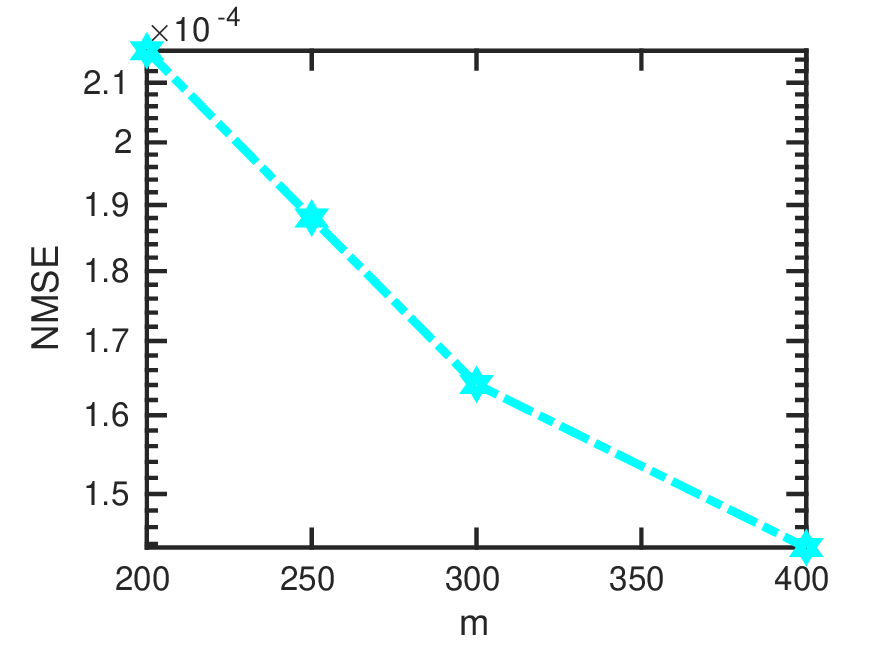}
	\caption{The effect of the number of time-varying threshold sequences $m$ in the input image reconstruction in terms of NMSE.
	}

	\label{figure_2}
\end{figure}

\section{Numerical Investigation}
\label{NUM}

To evaluate the performance of our reconstruction method, we employ the example of the "Grape" image (pet's name) illustrated in Fig.~\ref{figure_1}a. 
This image was obtained by the downsampling operator to form a $256 \times 144$ image.
We make use of the separability of B-spline basis functions for reconstruction, employing row-wise processing for recovery, with parameters set to $h_1$ and $h_2$ equal to 1, and a maximum pixel value of 255. We also incorporate a parameter $\lambda$ set to 1. In the image recovery process, we utilize the UNO reconstruction algorithm outlined in Section~\ref{sec3}, employing parameters $N=3$, $l=2$, $T=0.005$, and $\boldsymbol{\uptau}^{(\ell)}\sim\mathcal{U}_{\left[-1,1\right]}$ for $\ell\in[m]$.
As shown in Fig.~\ref{figure_1}b, the resulting image exhibits high quality, with no discernible reconstruction artifacts visible to the naked eye. This result indicates that high quality HDR imaging is possible when our simple proposed sampling scheme is utilized. Fig.~\ref{figure_1}c illustrates the obtained UNO data for an arbitrary sequence of threshold within our proposed sampling architecture.
Fig.~\ref{figure_2} shows the effect of the number of time-varying sampling threshold sequences $m$ in the input image reconstruction accuracy. As can be observed, increasing the number of dithering sequences further enhances the quality of the reconstructed image. Note that this result is obtained by averaging over $1000$ experiments. Our recovery method exhibits empirical stability in the presence of noise, as demonstrated in \cite{leventhal2010randomized}, where it is shown that RKA remains robust under noisy conditions.

\section{Concluding Remarks}
\label{sec:4}

We extended the UNO sampling scheme from its original application in bandlimited input signals to non-bandlimited signals within spline spaces, particularly relevant in imaging applications like HDR imaging.
In the realm of UNO sampling for HDR imaging, where we aim to minimize information loss due to sensor saturation, we employ a two-step approach. We begin by applying the modulo operator to the input signal, followed by employing a one-bit quantizer on the modulo samples to capture one-bit data, which is later used in the reconstruction stage. For this one-bit signal reconstruction, we utilized the RKA algorithm and provided a comprehensive analysis of its convergence to the original modulo samples.
We also introduced a novel sampling theory tailored to our proposed scheme, which extends the sampling theory for unlimited sampling in the context of HDR imaging for the case of corrupted modulo samples. We established that for this sampling rate to be effective in UNO sampling, the error in the one-bit signal reconstruction process must satisfy a specific bound, for which we derive the minimum probability.

\bibliographystyle{IEEEtran}
\bibliography{references}

\end{document}

%% file: symbols.tex
\usepackage[dvipsnames]{xcolor}
\usepackage{amsmath}
\usepackage{amsthm}
\usepackage{dsfont}
\usepackage{thmtools}
\usepackage{amssymb}

\def\bGamma{\boldsymbol{\Gamma}}

\def\bOmega{\boldsymbol{\Omega}}

\def\mbb{\mathbf{b}}
\def\mbc{\mathbf{c}}

\def\mbe{\mathbf{e}}

\def\mbr{\mathbf{r}}

\def\mbx{\mathbf{x}}
\def\mby{\mathbf{y}}

\def\mbC{\mathbf{C}}

\newtheorem{theorem}{Theorem}
\newtheorem{proposition}{Proposition}

\theoremstyle{definition}

